\documentclass[twocolumn,aps,prb,superscriptaddress,a4paper,showpacs,showkeys,lengthcheck]{revtex4-1}

\usepackage{graphicx}
\usepackage{dcolumn}
\usepackage{bm}
\usepackage{type1cm}
\usepackage[normalem]{ulem}

\usepackage{amssymb,amsmath}
\usepackage{braket}

\usepackage[dvips]{color}

\newcommand{\new}[1]{#1}

\newcommand{\mm}[1]{\mbox{$#1$}}

\newcommand{\ms}{\mbox{$\mu_{\mathrm{spin}}$}}
\newcommand{\mstz}{\mbox{$\mu_{\mathrm{spin}}+7T_{z}$}}
\newcommand{\mo}{\mbox{$\mu_{\mathrm{orb}}$}}
\newcommand{\moms}{\mbox{$\mu_{\mathrm{orb}}/\mu_{\mathrm{spin}}$}}

\newcommand{\mB}{\mbox{$\mu_{B}$}}

\newcommand{\tz}{\mbox{$T_{z}$}}
\newcommand{\MM}{\mbox{$\bm{M}$}}

\newcommand{\ea}{{\it et al.}}

\begin{document}

\title{Trends in magnetism of free Rh clusters via relativistic
  ab-initio calculations}



\author{O. \surname{\v{S}ipr}} 
\email{sipr@fzu.cz}
\homepage{http://www.fzu.cz/~sipr} \affiliation{Institute of Physics
  of the ASCR v.~v.~i., Cukrovarnick\'{a}~10, CZ-162~53~Prague, Czech
  Republic }

\author{H. \surname{Ebert}} \affiliation{Universit\"{a}t M\"{u}nchen,
  Department Chemie, Butenandtstr.~5-13, D-81377~M\"{u}nchen, Germany}

\author{J. \surname{Min\'{a}r}} \affiliation{Universit\"{a}t
  M\"{u}nchen, Department Chemie, Butenandtstr.~5-13,
  D-81377~M\"{u}nchen, Germany} \affiliation{New Technologies Research
  Centre, University of West Bohemia, Pilsen, Czech Republic}


\begin{abstract}
A fully relativistic {\em ab-initio} study on free Rh clusters of
13--135 atoms is performed to identify general trends concerning their
magnetism and to check whether concepts which proved to be useful in
interpreting magnetism of 3$d$ metals are applicable to magnetism of
4$d$ systems.  We found that there is no systematic relation between
local magnetic moments and coordination numbers.  On the other hand,
the Stoner model appears well-suited both as a criterion for the onset
of magnetism and as a guide for the dependence of local magnetic
moments on the site-resolved density of states at the Fermi level.
Large orbital magnetic moments antiparallel to spin magnetic moments
were found for some sites.  The intra-atomic magnetic dipole \tz\ term
can be quite large at certain sites but as a whole it is unlikely to
affect the interpretation of x-ray magnetic circular dichroism
experiments based on the sum rules.
\end{abstract}



\maketitle


\section{Introduction}   \label{sec-intro}

Magnetism of systems with reduced dimensions is a vivid research area
both because of fundamental interest and because of technological
relevance.  To be able to design new devices based on magnetism of
clusters and thin films one should have an intuitive understanding for 
these, so that one can estimate beforehand which configurations
and compositions might be perspective or not.  Magnetism of Rh carries
a special appeal because it concerns a material which is magnetic only
in reduced dimensions and in which the spin orbit coupling (SOC) is
expected to play a more important role than in the magnetism of 3$d$
elements.  Indeed, Rh magnetism seems to be quite intriguing, with
results obtained so far being often controversial and not easy to
interpret.  Catalytic properties of Rh add yet another impetus for
studying low-dimensional Rh systems.

The interest in free Rh clusters was stimulated by a
Stern-Gerlach-type experiment on clusters of 9--100
atoms,\cite{CLAB94} where it was found that Rh clusters have magnetic
moments which decrease with cluster size down from 0.8~\mB\ in a
non-monotonous way and become strongly suppressed (possibly to zero)
for clusters of more than 60 atoms.  Another experimental input comes
from x-ray magnetic circular dichroism (XMCD) measurements of Sessi
\ea\cite{SKZ+10a} who studied quasi-free Rh clusters of few tens of
atoms in a Xe matrix on Ag(100) and found that the clusters are
magnetic, with an effective spin magnetic moment \ms\ of about
0.26~\mB\ and with the ratio between orbital and spin magnetic moments
\moms\ about 40\%.  Recently, spin and orbital moments induced in
paramagnetic Rh clusters of about 220 atoms embedded in a
Al$_{2}$O$_{3}$ matrix were determined also via XMCD\cite{BRW+12} and
it was found that the orbital moment \mo\ does not exceed 2\% of \ms.

For a deeper understanding of the magnetism of free Rh clusters and of
Rh magnetism in general it would be useful to identify general trends
which Rh-based systems obey.  This could be achieved by monitoring the
trends for a large set of related systems and comparing them with the
situation for (much better studied) 3$d$ elements.  Such a study
should be performed in a relativistic way because only then possible
effects of SOC can be properly included.  Also, experiments pose some
questions that call for fully relativistic calculations --- such as
the above mentioned issue of small \mo\ measured by XMCD for a set of
paramagnetic clusters embedded in Al$_{s}$O$_{3}$
(Ref.~\onlinecite{BRW+12}) in contrast to large \mo\ measured for
ferromagnetic clusters in a Xe matrix.\cite{SKZ+10a}

A considerable effort was devoted to theoretical research on free Rh
clusters recently.  Even though interesting and important results were
obtained, only few systematic studies for large sets of systems were
done so only an incomplete picture could have been gathered.  This is
even more true concerning relativistic properties such as orbital
magnetism.  Most {\em ab initio} studies focused on clusters of less
than 20 atoms and specifically on their geometries.  It turns out that
small clusters often tend to have non-compact
shapes.\cite{BOK+04,CC+04,AMG+06} The situation is still not clear
because different density functional theory (DFT) implementations
predict different structures and spin configurations.\cite{BZC+13}
Pseudopotential calculations seem to give rise to more open structures
than all-electron calculations.\cite{ABV+09} Structural studies for
large clusters were performed only within a model $d$-band
tight-binding (TB) Hamiltonian and it was found that the relaxation of
bond distances for fcc clusters of 13-165 atoms is non-uniform, with
some distances decreasing and some distances increasing.\cite{GDS+00}

It emerges that the existence of more competing configurations is
typical for Rh clusters and sometimes the configurations differ only
little in energies but considerably in magnetic
moments.\cite{JTK+94,Lee+97,KK+03,FMH05} It seems that a similar
situation arises regarding the suppression of magnetism for large
clusters, which occurs for sizes between 60 and 100 atoms according to
the experiment.\cite{CLAB94} E.g., model Hamiltonian calculations
found that for fcc clusters of 55 and 79 atoms magnetic solutions
exist but are higher in energy than non-magnetic
solutions.\cite{GDS+00} For clusters of 135 and 165 atoms, no magnetic
solutions were found.\cite{GDS+00} Pseudopotential calculations for
icosahedral clusters yield a finite moment for a 55-atoms cluster and
a zero moment for a 147-atoms cluster.\cite{KK+03} At the same time,
in order to explain their XMCD experiments, Barthem \ea\cite{BRW+12}
suggest that the Stoner criterion may be locally fulfilled at only
some atoms in large Rh clusters, giving rise to an inhomogeneous
magnetic state.

Connected with the question of size-dependence of magnetic moments is
the question of the dependence of local magnetic moments on the
coordination numbers.  For 3$d$ clusters (both free\cite{SKE+04} and
supported)\cite{MLZ+06,SBM+07,BSM+12} a distinct relation could be
identified in this respect. However, this need not be transferable to
4$d$ clusters and there are some hints that a relation of this kind
may be absent in Rh systems.\cite{ARM+02,BOK+04,AMG+06} However, no
systematic study about this has been performed so far.

The issue of orbital moments has not yet attracted much theoretical
attention.  Model $d$-band TB Hamiltonian calculations were performed
for few Rh clusters of up to 27 atoms.\cite{GVDP00,GDP+03} It was
found that \mo\ is typically 10--20\% of \ms.  The orbital moment
\mo\ is nearly always parallel to \ms\ and when it is antiparallel to
\ms, it is very small.  {\em Ab initio} calculations performed for Rh
clusters of 2--7 atoms lead to similar conclusions: spin and orbital
magnetic moments are always parallel to each other and the
\moms\ ratio is about 10\%, except for the smallest
clusters.\cite{YCK+12} In all these studies the SOC was added in a
second variation approach.  However, the SOC may play a more important
role in 4$d$ clusters and it is not clear whether it can be fully
accounted for in this way.  Recalling also that recent XMCD studies
led to opposite conclusions regarding the
\moms\ ratio,\cite{SKZ+10a,BRW+12} a more detailed study of orbital
magnetism of Rh clusters is desirable.

Concerning the interpretation of experiments, let us mention finally
that the XMCD sum rules do not provide a ``bare'' spin moment \ms\ but
only its combination with the magnetic dipole $T_{z}$ term, \mstz.  It
is conceivable that a disagreement between theory and experiment may
arise partly due to the effect of the $T_{z}$ term.\cite{BRW+12} Earlier
calculations for Rh$_{19}$ and Rh$_{43}$ clusters suggest that this
influence should not be crucial for free clusters\cite{SKZ+10a} but
this ought to be analyzed more deeply and for a large set of systems.

It turns out that despite the intense attention that has been devoted
to theoretical research on free Rh clusters, some important issues
remain to be solved.  One of them is whether there is any systematic
relation between local magnetic moments and coordination numbers as in
3$d$ clusters or whether such a relation is totally absent.  Linked to
this is a question whether the Stoner criterion can be applied
locally, i.e., whether it could be used to estimate local magnetic
moments in large clusters.  The relatively large SOC implies a
question to what extent will scalar-relativistic results differ from
fully-relativistic results.  How important is orbital magnetism in Rh
clusters, especially in large ones?  Is it possible to reconcile large
\moms\ ratios obtained via XMCD by Sessi \ea\cite{SKZ+10a} with small
\moms\ ratios obtained by Barthem \ea?\cite{BRW+12} What is the role
of the magnetic dipole term $T_{z}$?  Could it possibly be responsible
for the deviations between some theories and XMCD experiments?

To address these issues, we performed a fully-relativistic theoretical
study of clusters of 13--135 atoms and also of free-standing
monolayers (for comparison).  We found that some important intuitive
concepts which proved to be useful in magnetism of 3$d$ metals, such
as the relationship between magnetic moments and coordination numbers,
are not applicable to Rh clusters (and surfaces alike).  Some other
concepts, such as the Stoner model linking magnetic moments to DOS at
the Fermi level, remain to be valid and useful.


\section{Computational method}      \label{sec-comput}

The calculations were performed in an {\em ab-initio} way within the
spin density functional theory, relying on the local spin density
approximation (LSDA) 
\new{ as parametrized by Vosko \ea\cite{VWN80} }
The electronic structure is described fully 
relativistically by the corresponding Dirac equation.  The
computational approach is based on the multiple-scattering or
Korringa-Kohn-Rostoker formalism\cite{EKM11} as implemented in the
{\sc sprkkr} code.\cite{sprkkr-code} The electronic structure problem
is solved completely in real space.  The potentials were treated
within the atomic sphere approximation (ASA). For the angular momentum
expansion of the Green function, a cutoff of $\ell_{\text{max}}$=3 was
used.
\new{ Only collinear magnetic structures were considered in this
  study. }

The clusters were assumed to be spherical-like, containing 13--135
atoms, with fcc geometry as if cut from the bulk Rh crystal. They are
centered either around an atom (13, 19, 43, 55, 79, 87, and 135 atoms)
or around an interstitial empty site (14, 38, and 68 atoms).  No
structural relaxation was performed. The vacuum region was represented
by empty sites located at lattice points. The network of these empty
sites was built so that an empty site was included if it was a first-
or second-nearest neighbor to any cluster atom.

We also performed some comparative calculations for isolated or
free-standing (001), (110), and (111) monolayers and for the (001)
crystal surface.  These calculations were performed using supercell
(multilayer) geometries.  To simulate the crystal surface, we employed
a slab consisting of 17 layers of Rh atoms.  Successive Rh monolayers
or slabs were separated by 13--20~\AA\ of vacuum in the supercell.
The integration over the $\mathbf{k}$-points was done on a regular
mesh and for each of the systems the integration grid was chosen so
that its density corresponds to (at least) 15625 $\mathbf{k}$-points
in the full Brillouin zone of the bulk Rh crystal.  Otherwise the
calculations were done as for the free clusters.


\section{Results}   \label{sec-res}


\subsection{Magnetic moments}   \label{sec-moms}

\begin{table*}
\caption{Local magnetic moments for clusters of different sizes,
  resolved according to the distance $R$ from the cluster center.
  Each cluster is represented by one column.  The first line for each
  $R$ stands for $\mu_{\mathrm{spin}}$, the second for
  $\mu_{\mathrm{orb}}$ and the third for the ratio
  $\mu_{\mathrm{orb}}/\mu_{\mathrm{spin}}$ (shown only if
  $\mu_{\mathrm{spin}}$ is larger than 0.05~\mB).  The unit for $R$ is
  \AA, the unit for \ms\ and \mo\ is \mB. }
\label{tab-moms}
\begin{ruledtabular}
\begin{tabular}{lrrrrrrr}
  \multicolumn{1}{c}{$R$} & 
    \multicolumn{1}{c}{Rh$_{13}$} & 
    \multicolumn{1}{c}{Rh$_{14}$} & 
    \multicolumn{1}{c}{Rh$_{19}$} & 
    \multicolumn{1}{c}{Rh$_{38}$} & 
    \multicolumn{1}{c}{Rh$_{43}$} & 
    \multicolumn{1}{c}{Rh$_{55}$} & 
    \multicolumn{1}{c}{Rh$_{68}$} \\
\hline  
 0.00 & 
1.068  &    &  0.484 &     & -0.007 &  0.011 &     \\ 
      & 
-0.013 &   & -0.170 &    & -0.047 & 0.029 &   \\
      & 
-0.012 &   & -0.352 &    &   n/a &  n/a    &    \\  [0.5ex]
 1.90 & 
     & 1.193 &      &   0.457  &    &    &  0.01 \\
     & 
     &   0.032 &   & -0.046  &    &    &  0.01 \\
     & 
   &   0.027  &    &  -0.101  &    &    &  n/a  \\  [0.5ex]
 2.69 &
 1.452  &   & 0.864  &    &  0.099 &  0.296    &  \\
     & 
 0.118 &   &  0.138 &   &  -0.011 & -0.013  &   \\
    & 
 0.081 &   &  0.161 &   & -0.108  &  -0.044 &   \\  [0.5ex]
3.29  & 
     & 1.306 &   & 0.680  &    &     &   0.008  \\
      & 
    &  0.234 &   & -0.028   &    &    &  0.000 \\
     & 
    &   0.179   &    & -0.041  &    &   &  n/a   \\  [0.5ex]
3.80  & 
    &    &   0.632 &      & 0.495  & 0.847 &   \\
      & 
    &    &   0.100 &     & 0.028 & 0.050 &   \\
     & 
    &    &  0.160 &      & 0.056 & 0.059  &    \\  [0.5ex]
4.25  & 
  &  &  & 0.761  &   &   & 0.004    \\
      & 
  &  &  & 0.085 &   &   & 0.003    \\
     & 
  &  &  & 0.112 &   &   & n/a     \\  [0.5ex]
4.65  & 
   &  &  &  & 0.190  & 0.718   &    \\
      & 
   &  &  &  & 0.017 & 0.060  &   \\
     & 
   &  &  &  & 0.087 & 0.084  &    \\  [0.5ex]
5.38  & 
   &  &  &  &  & 0.254   &    \\
      & 
   &  &  &  &  & 0.035   &    \\
     & 
   &  &  &  &  & 0.139  &   \\  [0.5ex]
5.71  & 
   &  &  &  &  &  & 0.134  \\
      & 
   &  &  &  &  &  & 0.016  \\
     & 
   &  &  &  &  &  & 0.062  \\  
\end{tabular}
\end{ruledtabular}
\end{table*}

Local moments \ms, \mo\ and the ratio \moms\ for clusters of 13--68
atoms are shown in Tab.~\ref{tab-moms}.  Note that because of the
presence of magnetization and of the SOC, atoms belonging to the same
coordination shell need not be symmetry equivalent.  Each value in
Tab.~\ref{tab-moms} thus represents an average over all atoms located
at the same distance $R$ from the center of the cluster.  The
magnetization $\boldsymbol{M}$ is parallel to the (001) direction of
the parental fcc lattice.

No stable magnetic state was found for the 79-atoms and 87-atoms
clusters.  For the 135-atoms cluster a metastable magnetic state was
identified (see also Sec.~\ref{sec-ene} below), with sizable magnetic
moments only at the last two coordination spheres: the average
\ms\ (\mo) at the distance of 6.59~\AA\ is 0.091~\mB\ (0.017~\mB) and
average \ms\ (\mo) at the distance of 7.12~\AA\ is
0.271~\mB\ (0.025~\mB).

Concerning the angular-momentum decomposition, the $d$ component of
\ms\ comprises 90--95\% of its total value.  For \mo\ the dominance of
the $d$ component is even more pronounced.

\begin{figure*}
\includegraphics[viewport=0.5cm 0.5cm 18.5cm 6.5cm]{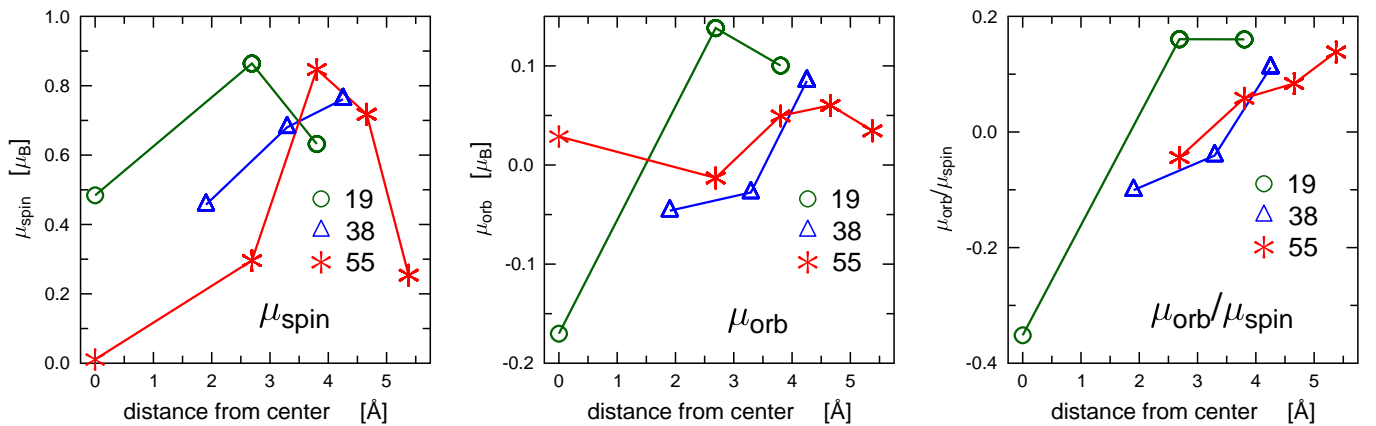}%
\caption{(Color online) Magnetic profiles for clusters of 19, 38, and
  55 atoms.  Left and middle panels show $\mu_{\mathrm{spin}}$ and
  $\mu_{\mathrm{orb}}$ averaged over all atoms at given distance from
  the center.  The right panel shows corresponding ratios
  $\mu_{\mathrm{orb}}$ and $\mu_{\mathrm{spin}}$ (the data point at
  $R$=0 for the 55-atoms cluster was omitted because the corresponding
  $\mu_{\mathrm{spin}}$ is very small).}
\label{fig-profile}
\end{figure*}

A better idea how the magnetic profiles look like may be obtained from
representative graphs in Fig.~\ref{fig-profile}.  It appears from all
the data that there is hardly any common trend in these profiles.
This is in contrast to the case of 3$d$ clusters where \ms\ and
\mo\ generally increase towards the surface for all cluster
sizes.\cite{SKE+04}


\subsection{Relation between local $\mu_{\mathrm{spin}}$ and the 
  coordination number} 

\label{sec-coord}

\begin{figure}
\includegraphics[viewport=0.5cm 0.5cm 9.0cm 7.0cm]{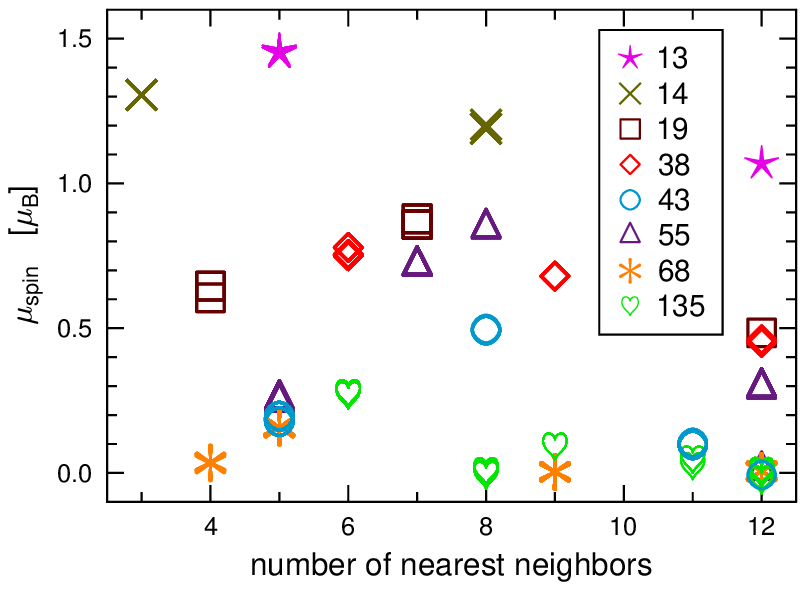}%
\caption{(Color online) Local $\mu_{\mathrm{spin}}$ as a function of
  the coordination number for atoms in clusters of 13--135 atoms.}
\label{fig-coordin}
\end{figure}

Our study covers a wide range of cluster sizes so it is possible to
perform a systematic search for a relation between local moments and
coordination numbers.  Earlier studies failed to find a common trend
like the one which had been identified before for the 3$d$ clusters
but in those studies always only a small range of cluster sizes was
explored.\cite{ARM+02,BOK+04,AMG+06} We display in
Fig.~\ref{fig-coordin} local \ms\ as a function of the coordination
number for magnetic clusters containing up to 135 atoms (including
thus also the metastable 135-atoms cluster).  It is evident from our
data that no common rule concerning all atoms can be drawn for free Rh
clusters.


\subsection{Average magnetic moments and magnetization energies} 
   \label{sec-ene}

A simple estimate of the strength of the magnetization can be obtained
from average magnetic moments and from magnetization energies
(differences of total energies per atom for a magnetic and a
non-magnetic state).  Furthermore, these quantities can be evaluated
using fully-relativistic calculations and using the
scalar-relativistic approximation,\cite{KH77} providing thus an idea
on the role of the SOC.  We summarized these results in
Tab.~\ref{tab-aver}.  Note that \mo\ is zero in the
scalar-relativistic approximation.

\begin{table}
\caption{Average magnetic moments in $\mu_{B}$ and magnetization
  energies per atom in mRy for clusters of 13--135 atoms. Apart from
  fully-relativistic results we display also \ms\ and $\Delta
  E_{\mathrm{mag}}$ obtained via scalar-relativistic calculations.}
\label{tab-aver}
\begin{ruledtabular}
\begin{tabular}{rrrrrr}
 \multicolumn{1}{c}{$N$}  & 
  \multicolumn{1}{c}{$\mu_{\mathrm{spin}}$}  & 
  \multicolumn{1}{c}{$\mu_{\mathrm{spin}}$}  & 
  \multicolumn{1}{c}{$\mu_{\mathrm{orb}}$}  & 
   \multicolumn{1}{c}{$\Delta E_{\mathrm{mag}}$} & 
   \multicolumn{1}{c}{$\Delta E_{\mathrm{mag}}$} \\ 
   & & \multicolumn{1}{c}{scalar}  &  &
   & \multicolumn{1}{c}{scalar} \\
\hline 
  13 &  1.423  &  1.604  &  0.108  & -11.74  & -13.82   \\
  14 &  1.258  &  1.433  &  0.147  &  -2.88  &  -3.55   \\
  19 &  0.771  &  0.823  &  0.110  &  -2.34  &  -2.63   \\
  38 &  0.696  &  0.785  &  0.040  &  -0.61  &  -1.47   \\
  43 &  0.202  &  0.254  &  0.009  &  -0.15  &  -0.22   \\
  55 &  0.526  &  0.563  &  0.037  &  -0.66  &  -1.15   \\
  68 &  0.062  &  0.088  &  0.009  &   0.00  &  -0.06   \\
 135 &  0.107  &  0.216  &  0.007  &   0.01  &   0.12   \\
\end{tabular}
\end{ruledtabular}
\end{table}

Average magnetic moments as well as magnetization energies generally
decrease with cluster size, albeit non-monotonously.  For the 68-atoms
cluster the magnetic and non-magnetic states are practically
degenerate, whereas for the 135-atoms cluster the ground state is
non-magnetic.  For a better understanding of the $\Delta
E_{\mathrm{mag}}$ values, recall that corresponding values for bulk
3$d$ metals are 31.2~mRy, 29.8~mRy, and 3.6~mRy for Fe, Co, and Ni,
respectively.  Our results indicate that for large Rh clusters
metastable isomers with different magnetic configurations should be
expected, similarly as it was found for small Rh
clusters.\cite{JTK+94,Lee+97,KK+03,FMH05}

One can see from Tab.~\ref{tab-aver} that the scalar-relativistic
approximation overestimates \ms\ typically by about 10\% (except for
large clusters and/or small \ms\ where the relative difference is
larger).  In line with this, scalar-relativistic calculations yield
larger magnetization energies.


\subsection{Stoner criterion applied locally} 

\label{sec-stoner}
 
\begin{figure}
\includegraphics[viewport=0.5cm 0.5cm 9.0cm 7.0cm]{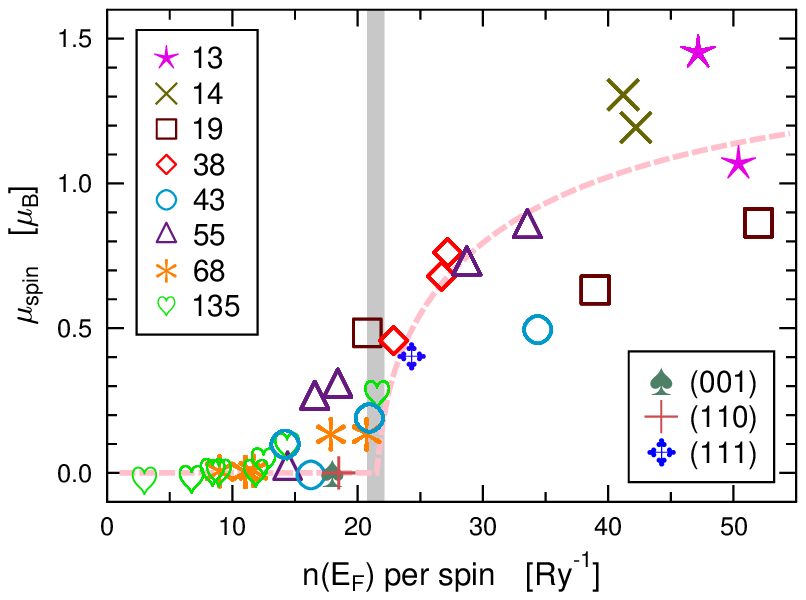}%
\caption{(Color online) Dependence of $\mu_{\mathrm{spin}}$ on the
  local density of states $n(E_{F})$ calculated for non-magnetic
  clusters. Corresponding data for free-standing monolayers are
  included as well. The vertical band highlights the critical value of
  $n(E_{F})$ when the Stoner product \mm{I_{s}n(E_{F})}\ becomes
  larger than one.  The dashed line is a fit to the Stoner model in
  Eq.~(\ref{eq-fit}).  }
\label{fig-stoner}
\end{figure}

When debating about magnetism of elements such as Ru, Rh, Pd, it is
often argued that these metals could become magnetic in lower
dimensions because then their density of states (DOS) at the Fermi
energy $n(E_{F})$ might increase due to band-narrowing, meaning that
the Stoner criterion
\begin{equation}
I_{s} \, n(E_{F}) \: > \: 1 \quad , 
\label{eq-stoner}
\end{equation}
with $I_{s}$ being the Stoner integral, would be satisfied.  It would
be interesting to see to what extent does this criterion really govern
magnetism of Rh.  It would be instructive to look not only on the
clusters but on surfaces and free-standing monolayers as well.

The Stoner exchange integral $I_{s}$ is considered to be a
quasi-atomic property, depending only very little on environmental
effects such as bonding etc.  We used the {\sc sprkkr} code to
evaluate $I_{s}$ for bulk Rh, for the (001) crystal surface and for
free-standing (001), (110), and (111) monolayers.\cite{Gun76} The
results are shown in Tab.~\ref{tab-stoner}.  One can see that indeed
the exchange integral $I_{s}$ varies only little from one Rh system to
another.

\begin{table}
\caption{Stoner exchange integral $I_{s}$, DOS per spin channel at
  $E_{F}$ in a non-magnetic state, Stoner product
  \mm{I_{s}\,n(E_{F})}\ and magnetic moment for
  two-dimensional Rh systems.}
\label{tab-stoner}
\begin{ruledtabular}
\begin{tabular}{lrrrr}
 \multicolumn{1}{c}{\mbox{system}}  & 
  \multicolumn{1}{c}{$I_{s}$}  & 
  \multicolumn{1}{c}{$n(E_{F})$}  & 
  \multicolumn{1}{c}{$I_{s}\,n(E_{F})$}  & 
   \multicolumn{1}{c}{$\mu_{\mathrm{spin}}$}  \\
\hline 
  &  \multicolumn{1}{c}{\mbox{mRy}}  & 
  \multicolumn{1}{c}{\mbox{Ry$^{-1}$}}  & 
    & 
   \multicolumn{1}{c}{\mbox{\mB}}  \\
\hline 
 (100) monolayer & 46.7 &  18.0  &  0.84  & 0.00  \\
 (110) monolayer & 45.5 &  18.5  &  0.84  & 0.00  \\
 (111) monolayer & 45.2 &  24.3  &  1.10  & 0.40  \\
 (100) surface   & 44.9 &  11.4  &  0.51  & 0.00  \\
 bulk            & 48.2 &   9.9  &  0.48  & 0.00  \\
\end{tabular}
\end{ruledtabular}
\end{table}

Eq.~(\ref{eq-stoner}) contains the DOS per spin channel $n(E_{F})$ for
the non-magnetic state.  Strictly speaking, $n(E)$ is a smooth
function of $E$ only if the energy levels form a continuous spectrum,
which is the case for infinite solids but not for finite clusters,
where the DOS is a sum of $\delta$-functions,
$n(E)=\sum_{j}\delta(E-E_{j})$. However, for already relatively small
clusters of 10--20 atoms one approaches a quasicontinuum regime, with
an energy level distribution that resembles that of a bulk
crystal.\cite{YJS+81,LCD84} One can construct an approximate DOS by
broadening the energy levels by a Gaussian\cite{YJS+81,LCD84} or
Lorentzian.\cite{LKJ+91} In our Green function formalism the same is
achieved by adding to $E$ a small imaginary part \mm{\mathrm{Im}E}, which is
equivalent to broadening the levels by a Lorentzian with full width at
half maximum of \mm{2\mathrm{Im}E}.  In this way we can evaluate $n(E_{F})$
for every atom in all the clusters we investigate.  If we choose
\new{ \mm{\mathrm{Im}E}=3~mRy, } 
we obtain a graph for the dependence of the local
\ms\ on the corresponding $n(E_{F})$ shown in
Fig.~\ref{fig-stoner}. The thick vertical gray line in
Fig.~\ref{fig-stoner} highlights the ``critical DOS'' which separates
magnetic systems from non-magnetic systems according to
Eq.~(\ref{eq-stoner}).

One can see that the Stoner criterion (\ref{eq-stoner}) can be used
with a good accuracy to predict whether a particular Rh atom will be
magnetic or not.  To be more precise, it follows from
Fig.~\ref{fig-stoner} that the criterion (\ref{eq-stoner}) is a bit
too strict --- one can get magnetism also for atoms where $n(E_{F})$
is by 10--20\% lower than the critical value.  If the Stoner model for
the dependence of \ms\ on $n(E_{F})$ is adopted,\cite{Sol10}
\begin{align}
\mu_{\mathrm{spin}} \: \sim \: \sqrt{ 1 \, - \,
\frac{1}{I_{\text{s}} \, n(E_{F})} }
 & \qquad \Leftrightarrow \quad n(E_{F}) > \frac{1}{I_{\text{s}}}
\;\; ,  \raisetag{-2\baselineskip}\label{eq-fit}  \\
\mu_{\mathrm{spin}} \: = \: 0
& \qquad  \Leftrightarrow \quad n(E_{F}) < \frac{1}{I_{\text{s}}}
 \notag
\end{align}
(see the Fig.~\ref{fig-stoner}), the best fit is obtained for
$I_{\text{s}}$=46.1~mRy, in very good agreement with the directly
computed values shown in Tab.~\ref{tab-stoner}.  So it turns out that
the Stoner model 
\new{ can serve as a guide to } 
Rh magnetism, even
concerning its local aspects.  We verified that this does not depend
on the particular value of the DOS broadening we used.  In fact graphs
similar to Fig.~\ref{fig-stoner} can be obtained if \mm{\mathrm{Im}E} is
varied within a reasonable interval 
\new{ 1--10~mRy. }

Corresponding data for bulk and two-dimensional Rh systems are shown
in Tab.~\ref{tab-stoner}.  The only two-dimensional system which is
magnetic is the (111) free-standing monolayer and it is also the only
one which satisfies the criterion given by Eq.~(\ref{eq-stoner}).  We
included these data in Fig.~\ref{fig-stoner} to demonstrate that they
follow the same trend as the data for clusters.


\subsection{Anisotropy of $\mu_{\mathrm{orb}}$ and $T_{z}$} 

\label{sec-tz}

\begin{figure*}
\includegraphics[viewport=0.4cm 0.5cm 12.5cm 6.5cm]{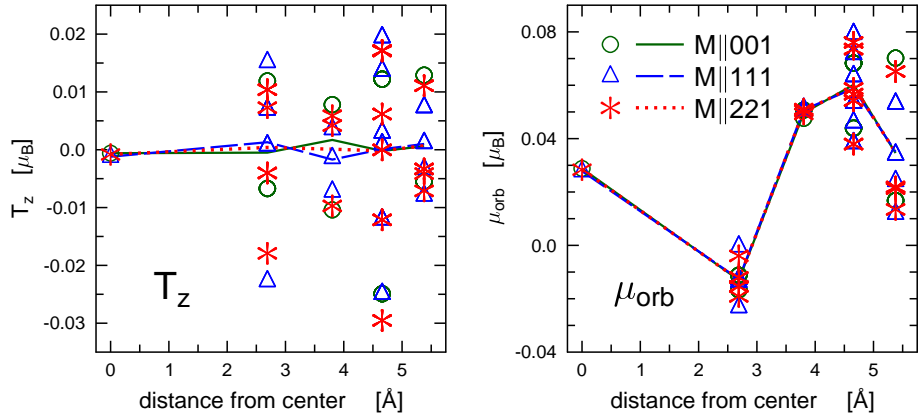}%
\caption{(Color online) Dependence of the $d$ component of the $T_{z}$
  term and of $\mu_{\mathrm{orb}}$ on the magnetization direction for
  the 55-atoms cluster.  Markers show values associated with
  individual atoms, lines show values averaged over all atoms at the
  same distance. The orientation of the magnetization $\bm{M}$ with
  respect to the underlying fcc lattice is specified in the legend. }
\label{fig-aniso}
\end{figure*}

The relatively strong SOC for Rh implies that \mo\ could play a more
important role than in the case of 3$d$ clusters.  Its effects are
also accessible to experiment via XMCD; partially conflicting result
were obtained in this respect.\cite{SKZ+10a,BRW+12} Linked to the use
of the XMCD is also the magnetic dipole term because the XMCD sum
rules deal not with \ms\ alone but only with a combination
\mm{\mu_{\text{spin}} + 7T_{z}}, where $T_{z}$ is the expectation
value of the intra-atomic dipole operator for the valence $d$
electrons (assuming that the $z$ axis is chosen parallel to $\bm{M}$).
Both \mo\ and $T_{z}$ depend on the magnetization
direction.\cite{SK95a,DL96,Sto99} This dependence was studied for 3$d$
free clusters\cite{SKE+04} and adatoms\cite{SBE+13} in the past.  The
$T_{z}$ term can significantly affect the perceived dependence of
\ms\ on the cluster size\cite{SME09b} and it was suggested that this
might occur also for 4$d$ clusters.\cite{YCK+12}

We calculated \mo\ and \tz\ for several magnetization directions, for
all the cluster sizes we explore here.  As a representative example we
select the 55-atoms cluster with \MM\ parallel to the [001], [111],
and [221] crystallographic directions.  The results are shown in
Fig.~\ref{fig-aniso}.  One notices immediately the effect of the
decrease of the symmetry due to the magnetization and the SOC: there
is a significant spread in the \tz\ and \mo\ values for atoms which
are at the same distance from the center.  There is also a significant
dependence on the direction of \MM\ for \tz\ and \mo\ of the
individual atoms.

When looking at the \tz\ term, we can see that as concerns individual
atoms, \tz\ could indeed be the cause of a significant difference
between \ms\ and \mstz; the magnitude of 7\tz\ is sometimes comparable
to the corresponding \ms\ for the Rh$_{55}$ cluster (see also
Tab.~\ref{tab-moms}).  However, the averaged or ``cumulative'' values
of \tz\ are very small, as evidenced by the lines in the left panel of
Fig.~\ref{fig-aniso}.  This is apparently linked to the fact that 
our clusters are spherical-like and therefore averaging over atoms for
a fixed magnetization direction is approximately equivalent to
averaging over magnetization directions for a fixed atom.  Such an
averaging should yield a small value for \tz\ due to the approximate
relation \mm{T_{x}+T_{y}+T_{z}=0}\ (valid exactly in the absence of
the SOC).\cite{SK95a} Therefore, as the measured XMCD always reflects
all atoms in the cluster, it seems that the \tz\ term cannot affect
the interpretation of XMCD experiments for free Rh clusters.  At least
--- not as long as they remain approximately spherical.

Concerning the orbital moments (right panel of Fig.~\ref{fig-aniso}),
again we can see a significant anisotropy when focusing on individual
atoms but hardly any anisotropy when focusing on the average values
(similarly as for free Fe clusters).\cite{SKE+04}
This apparently reflects the fact that the geometry of the clusters is
cubic.

\section{Discussion}

Our goal was to search for common trends governing magnetism of Rh
clusters, especially those with more than about 20~atoms.  We found
that there is no systematic link between magnetic moments and
coordination numbers. On the other hand, systematic trends can be
identified in the dependence of \ms\ on $n(E_{F})$.  Metastable
magnetic states exist for large clusters.  The influence of the SOC on
magnetism of Rh is not negligible but not crucial either.

The lack of a systematic relation between local magnetic moments and
coordination numbers is perhaps the most striking feature of magnetism
of Rh in comparison with magnetism of 3$d$ metals.  It means that
intuitive arguments which are routinely used when analyzing magnetism
of 3$d$ metals cannot be applied.  This can be illustrated by the case
of free-standing Rh monolayers of different crystallographic types
(see Tab.~\ref{tab-stoner}) where the only one which is magnetic is
the one with the {\em largest} coordination number.  
\new{ It is interesting to note that another kind of unusual trend in
  Rh magnetism was observed before, namely, the increase of magnetic
  moments of Rh adatoms on Ag(001) with decreasing interatomic
  distance.\cite{SHR+98} Unusual magnetic behaviour of Rh can thus be
  expected in a variety of environments.  }

We conjecture that the ``anomalous'' behavior of Rh 
\new{ observed here } 
is linked to the fact that 4$d$
electrons are less localized than 3$d$ electrons and thus the
dominance of the nearest-neighbors influence over the influence of
more distant neighbors is much smaller for the 4$d$ metals than for
the 3$d$ metals.  To be quantitative, for metallic Co, which is a 3$d$
analogue of Rh, the center of mass of the valence $d$ wave function
defined as \mm{r_{3d}=\left| \bra{\psi_{3d}} r \ket{\psi_{3d}} \right|
} is at about 0.52~\AA, which presents 21\% of the bulk interatomic
distance. (A $\psi_{3d}(\bm{r})$ wave function obtained for a
representative energy corresponding to the highest majority-spin
states DOS was used.) This should be compared to the $d$ wave function
for metallic Rh for which one gets $r_{4d}\approx0.78$~\AA, which
presents 31\% of the bulk interatomic distance. Our results also
complement arguments of Mavropoulos \ea\cite{MLZ+06} concerning
magnetic moments of supported 3$d$ and 4$d$ clusters.

The Stoner model appears well-suited for describing magnetism of 4$d$
metals.  It proves to be useful as a criterion for the onset of
magnetism, as it was demonstrated recently for Rh monolayers on noble
metals surfaces.\cite{GCB+03} However, it serves also as a guide for
the dependence of the magnitude of local magnetic moments on the DOS
at the Fermi level.  Here we would like to mention that a relationship
between local \ms\ and $n(E_{F})$ was observed also for thin Pd
slabs,\cite{HLW+07} again underlining the relevance of the Stoner
model for magnetism of 4$d$ systems.  This supports the argumentation
of Barthem \ea\cite{BRW+12} that in large Rh clusters some atoms may
be magnetic and others non-magnetic because $n(E_{F})$ may differ a
lot from one atom to another.

Even though the SOC does not play a prominent role in Rh magnetism,
there may be situations where it is quite important.  This is
illustrated by the case of the central atom of the Rh$_{19}$ cluster
where \mo\ reaches 35\% of \ms\ and is oriented {\em antiparallel} to
\ms\ (see Tab.~\ref{tab-moms}).  Antiparallel orientation of \mo\ can
be observed also for the Rh$_{38}$ cluster.  This finding is
interesting because an antiparallel orientation of \mo\ for some atoms
might explain why only very small averaged orbital moments were found for
embedded Rh clusters via the XMCD measurements of Barthem
\ea\cite{BRW+12}

\begin{table}
\caption{Magnetic moments in \mB\ from our fully-relativistic {\em
    ab-initio} calculations compared to results obtained via a model
  $d$-band TB Hamiltonian with an additional SOC term.\cite{GVDP00}
  Results for the Rh$_{13}$ and Rh$_{19}$ clusters are resolved
  according to distances of atoms from the center of the cluster, as
  indicated by the indices in the parantheses in the first column.
  Two different values of \mo\ are often present because magnetization
  together with the SOC decreases the symmetry in a fully relativistic
  treament.  }
\label{tab-compar}
\begin{ruledtabular}
\begin{tabular}{llcr}
 \multicolumn{1}{c}{shell}  &   & 
  \multicolumn{1}{c}{this work}  & 
  \multicolumn{1}{c}{model Hamiltonian\protect\cite{GVDP00}}  \\
\hline 
  Rh$_{13}$ (1)  &  \ms  &  1.07        &  1.19   \\
  Rh$_{13}$ (1)  &  \mo  &  -0.01       &  -0.01  \\  [0.5ex]
  Rh$_{13}$ (2)  &  \ms  &  1.45        &  1.62   \\
  Rh$_{13}$ (2)  &  \mo  &  0.10, 0.15  &  0.30   \\  [1.0ex]
  Rh$_{19}$ (1)  &  \ms  &  0.48        &  1.21   \\
  Rh$_{19}$ (1)  &  \mo  &  -0.17       &  0.13   \\  [0.5ex]
  Rh$_{19}$ (2)  &  \ms  &  0.86        &  1.11   \\
  Rh$_{19}$ (2)  &  \mo  &  0.17, 0.07  &  0.14   \\  [0.5ex]
  Rh$_{19}$ (3)  &  \ms  &  0.63        &  0.70   \\
  Rh$_{19}$ (3)  &  \mo  &  0.14, 0.08  &  0.16   \\
\end{tabular}
\end{ruledtabular}
\end{table}

A large antiparallel \mo\ for some Rh clusters was obtained by our
fully-relativistic {\em ab-initio} calculations but not by earlier
model $d$-band TB Hamiltonian calculations where the SOC was added
within the second variation approach.\cite{GVDP00} A detailed
comparison between our results and the model Hamiltonian results is
presented in Tab.~\ref{tab-compar} for Rh$_{13}$ and Rh$_{19}$
clusters (moments for the same geometries are shown).  One can see
that there is a general agreement between both calculations but
differences occur for some atoms.

All our calculations were done for bulk-like, i.e., unrelaxed
geometries.  Changing the interatomic distances would of course affect
the calculated \ms\ and \mo.  {\em Ab-initio} calculations for
clusters of 2--19 atoms found that bond-length contraction of 4--8~\%
may lead to large changes of magnetic moments, in particular for a
4-atoms cluster and a 10-atoms cluster.\cite{JTK+94} However, the bond
length changes are less important for large clusters --- in fact,
model Hamiltonian calculations suggest a bond-length contraction of
about 4~\% for clusters of 8--13 atoms but only 1--3~\% for clusters
of 55 and 147 atoms.\cite{BGS+00} Another study based on
semi-empirical potential and model $spd$ TB Hamiltonians found that
the average nearest-neighbor distance is close to the bulk-like
distance for clusters of more than about 20 atoms.\cite{ARM+02}
Let us also mention a model Hamiltonian study which found that
using equilibrium bond lengths instead of bulk bond lengths changes
\ms\ and \mo\ by about 50~\% for a cluster of 13 atoms while for
clusters of 19 and 27 atoms the bond relaxation decreases \ms\ only by
about 10\% and \mo\ by about 20\%.\cite{GVDP00} 
\new{ The fcc growth mode adopted here is plausible in the light of
  the {\em ab initio} study of Futschek \ea\cite{FMH05} who found that
  clusters of more than about ten atoms adopt ground state geometries 
  that can be considered as fragments of the fcc crystal structure. }
We conclude that while
small clusters of less than $\sim$10 atoms are quite sensitive to
structural relaxation, results for large Rh clusters are not so
dependent on whether the structural relaxation has been performed or
not.

When compared with experiment on free Rh clusters,\cite{CLAB94} our
calculation gives higher average magnetic moments in the 13--55 atoms
size range.  A $d$ band TB model Hamiltonian study of fcc
clusters\cite{GDS+00} gave similar values as our study while an {\em
  ab initio} pseudopotential-based calculations of icosahedral
clusters\cite{KK+03} led to smaller values, in a better agreement with
the experiment.  However, one should be cautious about drawing
conclusions, say, about preference of the icosahedral geometry over
the fcc geometry just from this: in all the theoretical studies, only
few cluster sizes were considered beyond the 20-atoms limit and the
dependence of the average magnetic moment on the cluster size displays
a quasi-oscillatory behavior, so one cannot be sure how representative
the results are.  One should also take into account that in the case
of small clusters, where more theoretical studies we performed, the
results often differ substantially from one study to
another.\cite{BOK+04,CC+04,AMG+06,BZC+13,ABV+09,KK+03,FMH05} It turns
out that accurate calculations which could be directly compared to
experiment are very difficult to perform because the systems are quite
complex and there are many factors to control.  Therefore, it is
useful to perform a study like the current one which focuses on the
analysis of general trends and rules which the Rh cluster should obey,
to improve our intuitive understanding.

\section{Conclusions}

Some important intuitive concepts which proved to be useful in
interpreting the magnetism of 3$d$ metals are not applicable to
magnetism of 4$d$ systems such as Rh clusters.  In particular there is
no systematic relation between local magnetic moments and coordination
numbers.  On the other hand, the Stoner model describes even local
aspect of Rh magnetism quite well; atoms for which the DOS in the
non-magnetic state $n(E_{F})$ is more than 20 states per spin per Ry
will be magnetic.  

Fully-relativistic calculations indicate that there can be large
\mo\ antiparallel to \ms\ for some atoms.  The intra-atomic \tz\ term
can be quite large at certain sites but as a whole it is unlikely to
affect the interpretation of XMCD experiments based on the sum rules.
For clusters in the $\sim$100~atoms size range it is likely that there
will be metastable magnetic configurations, with excitation energies
less than 0.1~mRy per atom.


\begin{acknowledgments}
Financial support by the Grant Agency of the Czech Republic within the
project 108/11/0853 and by the Deutsche Forschungsgemeinschaf within
the project SFB 689 is gratefully acknowledged.  J.M.\ also
acknowledges support by the CENTEM project CZ.1.05/2.1.00/03.0088,
co-funded by the ERDF as part of the Ministry of Education, Youth and
Sports OP RDI programme.
\end{acknowledgments}





%

\end{document}